\documentclass[prl,twocolumn,preprintnumbers,amsmath,amssymb]{revtex4}

\usepackage{bm,curves}
\usepackage[colorlinks=true,linkcolor=blue]{hyperref} 
\usepackage{color}
\usepackage{dcolumn}
\usepackage{ulem}

\begin{document}

\title{Reply to Comment on `Light deflection by Damour-Solodukhin wormholes and Gauss-Bonnet theorem'
by Amrita Bhattacharya and Kh. Karimov Ramis}

\author{Ali \"{O}vg\"{u}n}
\email{ali.ovgun@pucv.cl}
\homepage{http://www.aovgun.com}
\affiliation{Instituto de F\'{\i}sica, Pontificia Universidad Cat\'olica de
Valpara\'{\i}so, Casilla 4950, Valpara\'{\i}so, Chile.}

\affiliation{Physics Department, Arts and Sciences Faculty, Eastern Mediterranean University, Famagusta, North Cyprus via Mersin 10, Turkey.}

\date{\today}

\begin{abstract}
We reply to the  Comment of Bhattacharya and Karimov on our paper 
``Light deflection by Damour-Solodukhin wormholes and Gauss-Bonnet theorem'' [arXiv:1811.00768 [gr-qc]].
We address a number of incorrect claims made about our calculations and methodology. We show below that the claims of BK are incorrect, and that there are no problems with the results of  \"{O}vg\"{u}n's paper or its implications. 
\end{abstract}
\maketitle

In a recent Comment by, Bhattacharya and Karimov (BK) \cite{Bhattacharya:2018leh} criticize
our analysis \cite{Ovgun:2018fnk} of light deflection by Damour-Solodukhin wormholes and Gauss-Bonnet theorem. The text and calculations of our paper already contain a thorough discussion of all these points, but to avoid any confusion we shall briefly address them here again.

\textbf{Reply:}
Given a background spacetime, the propagation of light in can be studied by means of various techniques. In my paper \cite{Ovgun:2018fnk}, we have applied a geometrical method introduced by Gibbons and Werner \cite{Gibbons:2008rj}, this method was later extended by Werner \cite{Werner:2012rc} to stationary spacetime metrics whose optical geometry is Finslerian, which uses the optical geometry whose geodesics are the spatial light rays and then extended by Ono, Ishihara and Asada (OIA)  \cite{Ishihara:2016vdc,Ono:2017pie}.

 Recently, it was investigated how this method may be used to calculate the deflection angle for finite distances \cite{Ishihara:2016vdc,Ono:2017pie,Arakida:2017hrm}. In this paper we have used Ono, Ishihara, and Asada (OIA) method and we find the following result for the deflection angle in the weak deflection approximation. This confirms that this approximation is indeed valid to recover the leading order terms in $M$, and $Ma$  but would need to be modified to correctly reproduce higher order terms such as this one.  As Werner mentioned in his paper \cite{Werner:2012rc}, this method is good to find the two leading terms of the asymptotic deflection angle of gravitational lensing \cite{Jusufi:2017lsl,Jusufi:2017hed,Jusufi:2017vta,Jusufi:2017vew,Jusufi:2017uhh,Ovgun:2018ran,Ovgun:2018prw,Ovgun:2018oxk,Ovgun:2018fte,Ovgun:2018tua,Ono:2018ybw,Sakalli:2017ewb,Jusufi:2018kmk,Jusufi:2017mav}. My calculations and results are correct for the deflection angle at expected leading terms in $M$ and $a$ of the asymptotic deflection angle which reduces to this form of Kerr black hole's deflection angle for $\lambda=1$

\begin{equation}
\alpha \approx \frac{4M}{b} \pm 	\frac{4Ma}{b^2}.	
\end{equation}

The authors presumably confuse the Gibbons and Werner approach.  Recently OIA have showed that it is possible to find deflection angle in  especially the numerical coefficients at the order of $M^2$ and $aM$ that agrees with the previous results \cite{Ono:2017pie}.

In conclusion, we show that the above claims are incorrect, and that there are no problems with the results of  \"{O}vg\"{u}n's paper or its implications. Note furthermore that the negative comments of BK that calculating the leading terms of the Kerr deflection angle are was not the goal of my paper. The statements made in the BK comments concern not my theory but another one where the difficulties they claim to exist appear \cite{Gibbons:2008rj,Werner:2012rc}.  We suggest to the authors of the BK to try to find leading order terms of deflection angles using the Gauss-bonnet theorem, which is the good problem.

\end{document}